\documentclass[twocolumn,aps,prb,english,twocolumn,epsfig,groupedaddress]{revtex4}
\usepackage{epsfig}
\usepackage{amsmath}
\usepackage[latin1]{inputenc}
\usepackage{latexsym}
\usepackage{ulem}
\usepackage{color}

\begin{document}
\title{Optical conductivity of Mn doped GaAs.}
\author{Georges Bouzerar$^{1,2}$ and Richard Bouzerar$^{1}$}
\affiliation{
$1.$ Institut N\'eel, CNRS, d\'epartement MCBT, 25 avenue des Martyrs, B.P. 166, 38042 Grenoble Cedex 09, France \\
$2.$ Jacobs University Bremen, School of Engineering and Science, Campus Ring 1, D-28759 Bremen, Germany \\
}
\date{\today}
\begin{abstract}
The optical conductivity in the III-V diluted magnetic semiconductor compound (Ga,Mn)As is investigated theoretically.
A direct comparison between calculations and available experimental data is provided. We demonstrate that our model study is able to reproduce both qualitatively and quantitatively the measurements performed by Burch et al. and Singley et al.. It is found that the increase of the carrier density up to one hole per Mn leads to a red shift of the broad conductivity peak located at approximately 200 meV in optimally annealed samples. Our study demonstrates that the non perturbative treatment (beyond the valence band picture) is crucial to capture these features. Otherwise a blueshift and an incorrect amplitude would result. We have calculated the Drude weight (order parameter) and have established the metal-insulator phase diagram. It is shown that (i) Mn doped GaAs is indeed close to the metal-insulator transition and (ii) in both 5$\%$ and 7$\%$ doped samples, 20$\%$ of the carriers only are delocalized. Beyond the mobility edge, we have found that the optical mass is m$_{\rm opt}$ $\approx$ 2 m$_{e}$. Interesting new features for overdoped samples have been found. The overdoped regime could be experimentally realized by Zn codoping. Detailed discussions and carefull analysis are provided.
\end{abstract}
\maketitle

The possibility to use for spintronics purpose the spin degree of freedom in magnetic semiconductors has lead to a tremendous experimental and theoretical work. One of the main goals is to search for optimal candidates that exhibit high Curie temperatures (beyond room temperature).
Although it is well accepted that the presence of holes in III-V diluted magnetic semiconductors is responsible for the induced ferromagnetism, the particular case of (Ga,Mn)As remains still controversial concerning the nature of the states at the Fermi level. There are in fact several important points that are still controversial in the literature. Is the Fermi level inside the weakly perturbed valence band (VB scenario) or does (Ga,Mn)As exhibit a well defined impurity band ? Are the holes extended or localized ? Or is the situation similar to that of Mn doped II-VI as (Zn,Mn)Te ? For those latter compounds the Valence Band picture is appropriate. It seems that this could be at the origin of some of the controversies.
In our view the main question should be: In the case of (Ga,Mn)As, is a perturbative treatment (Valence Band picture) appropriate to explain the physics ? Or equivalently, is it necessary to treat disorder effects properly? From ab-initio based studies, one finds that (Ga,Mn)As clearly exhibits a preformed impurity band \cite{Sanvito02,Bergqvist03,Sandratskii04} that is responsible for (i) the non RKKY nature of the magnetic couplings and (ii) explains the measured high Curie temperatures (with respect to II-VI compounds). Additionally, it has been clearly demonstrated that RKKY couplings obtained within the VB picture can not explain the magnetism in III-V compounds \cite{richard-prb}. Indeed, the frustration effects would lead to either a spin glass phase or to extremely small Curie temperatures.  On the other hand, the non perturbative treatment of the appropriate model Hamiltonien captures the essential physics and lead to non RKKY couplings \cite{bouzerar-VJ07}, in agreement with ab initio studies. Note also that, below 1\% the impurity band is completely splitted from the valence band. Early resistivity measurements have clearly demonstrated that (Ga,Mn)As is close to the metal-insulator phase transition. Indeed, measurements in as grown samples often show an insulator behaviour at low temperature and a metallic one after annealing \cite{Matsukura98,Hayashi01,Potashnik01,Edmonds02}. Of course, the VB scenario is unable to capture this physics, namely the metal-insulator transition.
It has been shown, by Singley et al. and Burch et al. that the measured optical conductivity is also inconsistent with the perturbative Valence Band scenario \cite{Singley02,Singley03,Burch06}. The aim of this paper is to provide a non perturbative theoretical study able to reproduce and explain the experimental observations.

Ab-initio based studies have provided the most reliable tool, to allow for quantitative studies of the magnetic properties without any adjustable parameters. As an example, it was possible to study in great details and quantitatively the magnetic properties of Ga$_{1-x}$Mn$_x$As, both in the presence and absence of native (compensating) defects \cite{bergqvist04,gbouzerar05a,gbouzerar05b}.
However, dynamical transport study is not a simple task within the ab-initio approach.
The study of relevant minimal model is thus necessary. The model approach is especially suitable to understand the influence of a particular physical parameter and  may provide a support for finding new potential spintronic candidates. The V-J model treated non perturbatively was shown to capture qualitatively the magnetic properties of a whole family of III-V materials \cite{bouzerar-VJ07}. The present theoretical study is based on such a model. The one band V-J Hamiltonian reads,
\begin{eqnarray}
H=-\sum_{ij} t_{ij} c^{\dagger}_{i\sigma}c_{j\sigma} + \sum_{i} p_{i}J_{i} {\bf S}_{i}\cdot {\bf s}_{i} +\sum_{i\sigma} \epsilon_{i}p_{i}c^{\dagger}_{i\sigma}c_{i\sigma}
\label{Hamiltonian}
\end{eqnarray}
In the first term, t$_{ij}$$=$t if i and j are nearest neighbors, otherwise t$_{ij}$$=$0. The random variable $p_i=1$ if the site is occupied by an impurity, otherwise it is 0. ${\bf S}$$_{i}$ denotes the localized impurity spin at site i ($|S_{i}|=5/2$) and ${\bf s}_{i}$ the spin 1/2 operator of the carrier. J$_{i}$ is the p-d coupling (J) between itinerant carrier (p-states) and localized Mn spin . The last term results from the substitution of Ga$^{3+}$ by Mn$^{2+}$: $\epsilon_{i}=$V if the site is occupied by Mn, otherwise it is 0. This additional crucial term allows to adjust the position of the hybridized p-d states with respect to the top of the VB \cite{bouzerar-VJ07,bouzerar-VJ2010}. In the following, $x$ denotes the Mn concentration and p the hole density. In the absence of compensating defects (As anti-sites or Mn interstitials) $p=x$. In order to perform reliable calculations, the local coupling between impurity and carrier is treated in a non perturbative way and the disorder and dilution effects exactly (no effective medium approach). This treatment allows for the localization of the itinerant carriers which is expected to strongly affect both magnetic and transport properties. The procedure is as follow. For a given impurity concentration $x$ and a given disorder configuration, the calculations are done by exact diagonalization of the Hamiltonian (\ref{Hamiltonian}) in both spin sectors. This provides the set of eigenvalues and eigenstates $\{E^{\sigma,c}_{r}, |\Psi^{\sigma,c}_{r}\rangle\}$ where $\sigma=\uparrow,\downarrow$ and $r=1,2,....,N$, that are used to evaluate the optical conductivity $\sigma(\omega)$ and the Mn-Mn magnetic couplings. The superscript $c$ denotes the configuration of disorder, it will be omitted in the following. Note that, the calculations are performed at T$ = 0$ K assuming that the impurity spins are aligned.
\begin{figure}[htbp]
\includegraphics[width=8.0cm,angle=-0]{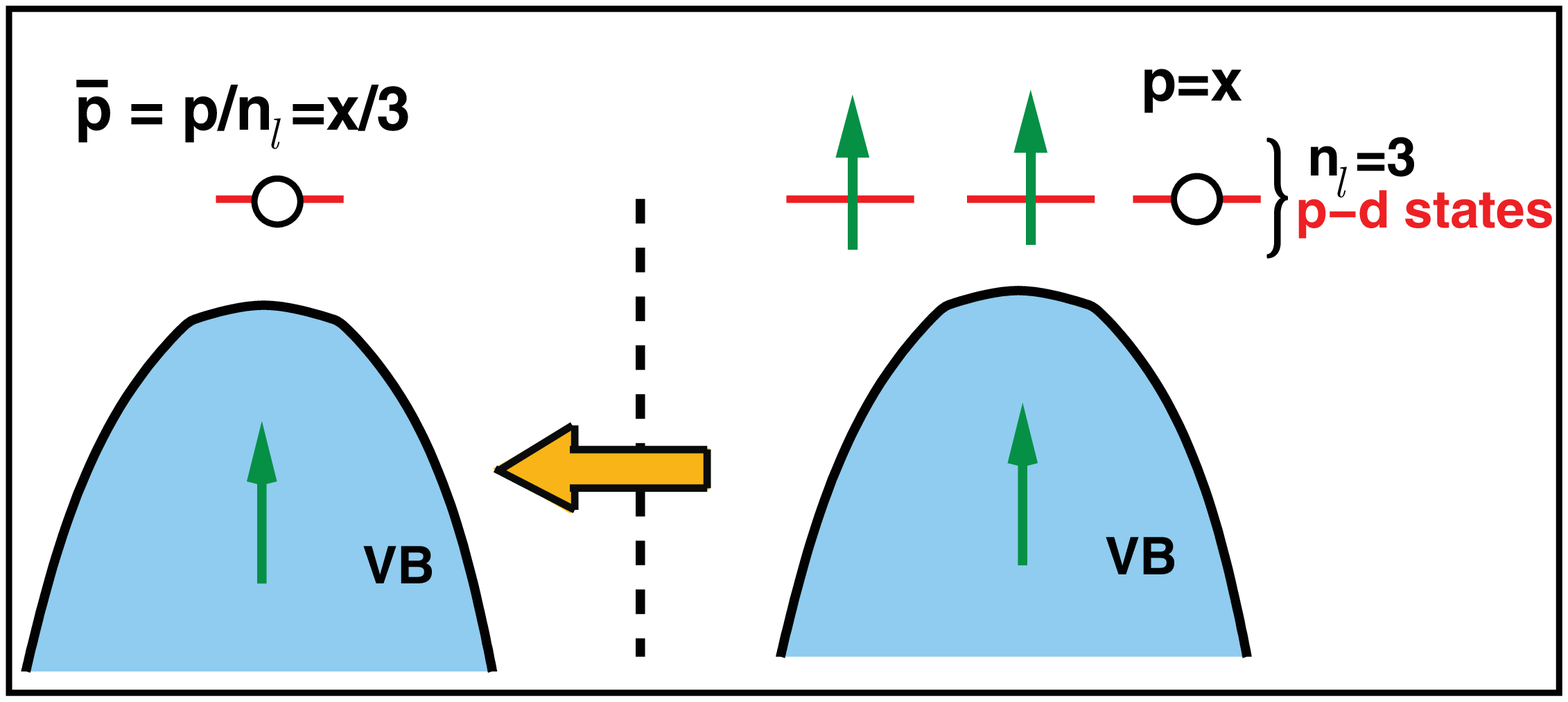}
\vspace{-1.30cm}
\caption{(Color online.) Illustration for uncompensated sample. In our model (left), one impurity brings one state above the top of the valence band. In reality (right), each Mn provides $n_{l}$ hybridized p-d states ($n_{l}=3$) and 1 hole/Mn. To be consistent, in the model the hole concentration is thus set to $\bar{p}=x/n_{l}$.
}
\label{Fig1}
\end{figure}

For simplicity, our calculations are performed on a simple cubic lattice
instead of the fcc lattice of Ga in the real GaAs compound. Thus, instead of 4 atoms/unit cell we have only one. Hence, the value of the lattice spacing that will be used is $a=\frac{a_{0}}{4^{1/3}}=3.55~ 10^{-10}$ m where $a_{0}=5.65 ~10^{-10}$ m is the lattice parameter of the Zinc-Blende GaAs material. Let us, shortly justify this choice. In these diluted compounds the carrier densities are very low, thus it is natural to expect the structure of the lattice to play no relevant role. As will be seen, this simplification will not have a significant effect on the results. This is also valid when the carrier are localized, as far as the localization length remains much larger than the lattice spacing.

Let us now discuss how the model parameters are fixed.
First, the hopping integral t is chosen according to the bandwith W of GaAs host. Thus, we use $t \approx 0.7~eV $ (W=12~t). JS is set to 3~eV since it is now well accepted that in (Ga,Mn)As J$\approx$1.2~eV \cite{Okabayashi98,Bhattacharjee00}. Note that J is also of the order 1~eV for other materials including II-VI compounds ((Zn,Mn)Te, (Cd,Mn)Te,..) \cite{Larson88,Gaj79,Aggarwal85,Heiman84,Twardowski83}.
The last parameter V (on-site energy) is as said above the crucial one. It actually explains the different nature of the Mn-Mn couplings and Curie temperatures values in Mn doped compounds such as (Zn,Mn)Te, (Ga,Mn)As and (Ga,Mn)N. The on-site energy V contains two contributions: (i) an electrostatic one due to the difference of charge between the host and substituted cation and (ii) another resulting from the Schrieffer-Wolff transformation of the Anderson Hamiltonian \cite{SW}. Indeed, in addition to $\sum_{i} p_{i} J_{i} {\bf S}_{i}\cdot {\bf s}_{i}$ term, this transformation leads to a spin independent scattering contribution which depends on the position of the d-levels with respect to the top of the valence band in the host material. V is set in order to reproduce the energy of the acceptor level E$_{b}$=110 meV in GaAs host \cite{Chapman67,Linnarsson97,Yakunin04,Kitchen05}. This leads to V$=$1.25~eV$=$1.8 t. Note that with this set of parameters, we were able to reproduce the whole variation of the Zeeman splitting as a function of $x$ obtained within first principle studies \cite{Wierzbowska04,Kudrnovsky(b)}. This splitting is defined by $\Delta (x,V)=E^{\uparrow}_{\rm max}-E^{\downarrow}_{\rm max}$, where $E^{\sigma}_{\rm max}$ is the largest eigenvalue in the $\sigma$-sector \cite{bouzerar-VJ2010}. For example, from ab-initio calculations \cite{Wierzbowska04,Kudrnovsky(b)}, $\Delta^{\rm ab-initio}=0.65$ eV for $x=0.05$, we obtain a value of 0.66 eV. Thus, from now on, the set of three parameters (t, J, V) is fixed and the calculations that will follow are done without any adjustable additional parameter. The calculations are performed for both 5\% and 7\% doped compounds in the ferromagnetic phase. Experimental and theoretical studies have shown that below 1\% no long range ferromagnetic order is possible (below the percolation threshold).  In the very dilute regime (below 1\%), the optical conductivity was recently studied in ref.\cite{moca09} within a very different theoretical approach. They perform some variational calculations on Mn-Mn molecule and extract the model parameters that are used to perform their calculations. 

Let us now explain in more details our theoretical method. The total optical conductivity reads $\sigma(\omega)=\sum_{\sigma}\sigma_{\sigma}(\omega)$ where $\sigma_{\sigma}(\omega)$ is given by the following expression \cite{Kohn,Millis,Scalapino},
\begin{eqnarray}
\sigma_{\sigma}(\omega)= D_{\sigma}\delta(\omega) + \sigma^{reg}_{\sigma}(\omega)
\label{optical}
\end{eqnarray}
$D_{\sigma}$ is the Drude weight ($\omega=0$ contribution) in the $\sigma$-sector and $\sigma^{reg}_{\sigma}(\omega)$ is the regular part of the optical conductivity. Note that the Drude weight is the order parameter for the metal-insulator phase transition: it is zero in the insulating phase and finite in the metallic one. In our calculations, we have used periodic boundary conditions in order to separate the finite frequency contribution from the dc part. In that case, the Kubo expression provides $\sigma^{reg}_{\sigma}(\omega)$ only. The Drude weight D is then obtained from the following sum-rule,
\begin{eqnarray}
D_{\sigma}= -\int_{0}^{\infty} \sigma^{reg}_{\sigma}(\omega)d\omega -  \frac{\sigma_{0}}{\hbar}  \frac{\langle\hat{K}_{\sigma}^{x}\rangle}{N}
\label{optical}
\end{eqnarray}
$\langle \hat{K}_{\sigma}^{x} \rangle$ is the average hole Kinetic energy in the $x$ direction and $\sigma$-sector,  N=L$^{3}$ is the total number of sites. 
and $\sigma_{0}=\frac{\pi e^{2}}{\hbar a}=21540 ~\Omega^{-1}\cdot $cm$^{-1}$. 

\begin{figure}[htbp]
\includegraphics[width=7.50cm,angle=0]{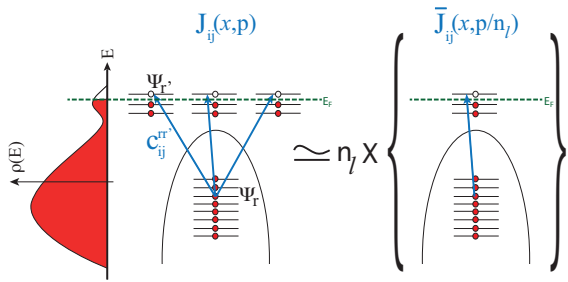}
\vspace{0cm}
\caption{(Color online.) Illustration for the procedure of calculation of the magnetic couplings between two Mn impurities located at site
i and j. The left part corresponds to the real compound in which each Mn$^{2+}$ provides 3 p-d hybridized states and one hole (this picture describes 3 Mn impurities in GaAs). The right side corresponds to the one band V-J model. The c$_{ij}^{rr'}$ are the matrix elements that enter the expression of the exchange given in eq. (\ref{eqct}).
 }
\label{Fig2}
\end{figure}

To allow for a direct comparison between available experimental data and theory, we have to take into account the following point: the substitution of Ga$^{3+}$ by Mn$^{2+}$ provides n$_{l}=3$ p-d states near the top of the valence band (see Fig.\ref{Fig1} for an illustration). These states are known as the Dangling Bond Hybrid states \cite{Zungerb}. As mentioned before, in the case of a single Mn impurity, the acceptor level energy is $E_{b}\approx 0.11 ~$eV. At finite concentration, an impurity band is formed and merges into the VB when $x$ becomes large enough $x \geq 0.010 $. Because Mn$^{2+}$ provides 1 hole per Mn, the resulting p-d hybridized states impurity band is thus 1/3 filled. On the other hand, in the one band V-J model, each impurity provides a single state.
Thus, to be consistent and coherent, if the (Ga,Mn)As compound contains a concentration $x$ of Mn and p
of holes the theoretical calculations should be done as follows. (1) The one band calculations are performed for the hole density $\bar{p}=\frac{p}{n_{l}}$ and (2) the theoretical optical conductivity that is compared to the experimental data is,
\begin{equation}
\sigma(\omega,p)=n_{l}\bar{\sigma}(\omega,p/n_{l})
 \label{conduc}
\end{equation}
where $\bar{\sigma}(\omega,p/n_{l})$ is that calculated within our one band model.
Similarly, the Curie temperature will be performed using the Mn-Mn magnetic couplings,
\begin{equation}
{J}_{{i},{j}}(x,p)=n_{l}\overline{{J}}_{{i},{j}}(x,\overline{p}=p/n_{l})
\label{Echangea}
\end{equation}
 where, $\bar{J}_{i,j}(x,p/n_{l})$ are those calculated within the one band V-J model. To be more specific these couplings are given by the following generalized susceptibility,
 \begin{equation}
 \overline{{J}}_{{i},{j}}(x,\overline{p})= -{\frac {1} {4\pi S^2}} \Im \int_{-\infty}^{E_F} Tr(\Sigma_{{i}}G_{{i},{j}}^{\uparrow}(\omega) \Sigma_{{j}} G_{{j},{i}}^{\downarrow}(\omega)) d\omega
\label{Echange}
\end{equation}
where the one particle retarded Green's functions G$_{i,j}^{\sigma}(\omega)=\langle i\sigma|\frac{1}{\omega -\hat{H} +i\epsilon} | j\sigma \rangle$. In the present case, the local exchange splitting energy is a constant $\Sigma_{i}=$J$_{pd}$S. An illustration of the procedure that provides the Mn-Mn exchanges is shown in Fig.\ref{Fig2}. Note that, this picture applies also for the optical conductivity calculation. In this case, the matrix elements c$_{ij}^{rr'}$ are replaced by those involving the current operator (see below).
 The previous equation can be rewritten as follows,
\begin{eqnarray}
\overline{{J}}_{{i},{j}}(x,\overline{p})=-\frac{1}{4 S^2}{\Sigma}_{i}{\Sigma}_{j} \sum_{r,r'}
c_{ij}^{rr'}\frac{n^{\uparrow}_{r}-n^{\downarrow}_{r'}}
{E^{\uparrow}_{r}-E^{\downarrow}_{r'}}
\label{eqct}
\end{eqnarray}
where n$^{\sigma}_{r}$ is the occupation number associated to the state $|\Psi^{\sigma}_{r} \rangle$. The
coefficients $c_{ij}^{rr'}=A_{ij,r}^{\uparrow}.A_{ji,r'}^{\downarrow}$, where $A_{ij,r}^{\sigma}=\langle i |\Psi_{\sigma}^{r}\rangle
\langle \Psi_{\sigma}^{r}|j \rangle$. The matrix elements $c_{ij}^{rr'}$ contains the relevant informations concerning the nature of the hole states, thus the nature of the magnetic couplings. In the VB picture r (resp. r') are replaced by the momentum $\overrightarrow{k}$ (resp. $\overrightarrow{ k'}$) and $c_{ij}^{rr'}=\frac{1}{N}e^{i(\overrightarrow{k}-\overrightarrow{k'})\cdot(\overrightarrow{r_{i}}-\overrightarrow{r_{j}})}$ leading to the standard RKKY couplings. The diagonalization of the resulting effective dilute Heisenberg Hamiltonian ${\cal H}_{Heis}= \sum_{ij}p_{i}p_{j} J_{i,j}(x,p) S_{i}\cdot S_{j}$ will provide the Curie temperature $T_{C}(x,p)$. This procedure will be done within the self-consistent local RPA (SC-LRPA) \cite{gbouzerar05a}. The self-consistent local RPA is a semi-analytical Green's function based method which is in particular able to capture quantitatively both the physics of localization and percolation. The disorder is treated in the real space (no effective medium) and the thermal/transverse fluctuations beyond standard Mean Field within the RPA. Comparisons with Monte Carlo treatment have demonstrated several times that the SC-LRPA is a reliable and very accurate tool (see for example \cite{gbouzerar05a,MC-LRPA}).

\begin{figure}[htbp]
\includegraphics[width=7.0cm,angle=-90]{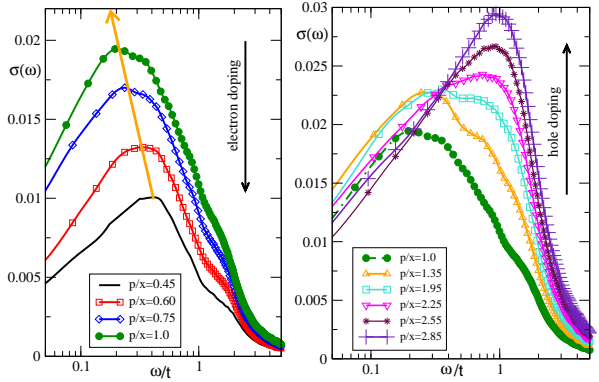}
\vspace{-1.3cm}
\caption{(Color online.) Regular part of the optical conductivity $\sigma(\omega)$ as a function of $\omega/t$ for $x=0.05$: (left) for 'underdoped' samples (electron doped) , e.g. from low carrier concentration to $p/x=1$ (well annealed sample) , (right) 'overdoped' samples (hole doped). The parameters are JS=4.3~t and V=1.8~t ((Ga,Mn)As set). The optical conductivity is given in units of $\sigma_{0}$ (see text). }
\label{Fig3}
\end{figure}

\begin{figure}[htbp]
\includegraphics[width=6.50cm,angle=0]{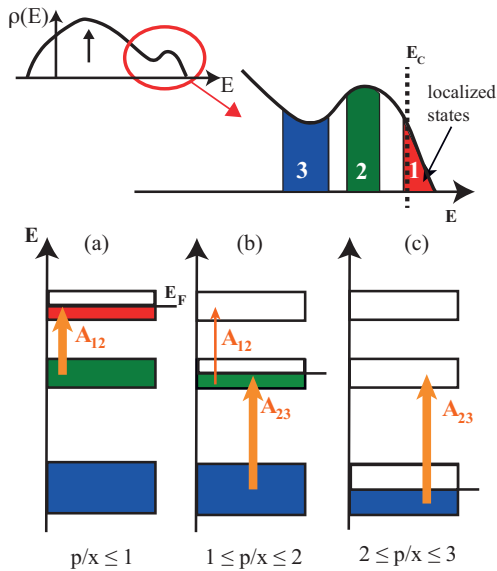}
\vspace{0cm}
\caption{(Color online.) The top of the figure is a sketch of the density of states for Mn doped GaAs: a compound with contains typically 5\% of Mn.
The zoom of the preformed impurity band is also shown. There are three relevant regions which provide the dominant transition matrix elements in the Kubo formula of the optical conductivity (see eq.(\ref{optical})). There are 3 relevant regions. Region (1) corresponds to the localized states (the mobility edge E$_C$ is indicated) and both Regions (2) and (3) to disordered extended states.
(a) For $p/x \le 1$ the dominant transition matrix element is A$_{12}$ between region (2) and region (1); (b) For $1 \le p/x \le 2$,  the matrix elements A$_{12}$ reduce and A$_{23}$ (between region 2 and 3) are now allowed and start to dominate; (c) for $2 \le p/x \le 3$, A$_{12}$ contributions are not possible anymore and the dominating process is A$_{23}$. Note that the width of the vertical arrows indicates the weight of the processes.}
\label{Fig4}
\end{figure}

The regular part of the optical conductivity $\overline{\sigma}_{\sigma}^{reg}(\omega,p/n_{l})$, is given by the Kubo expression which reads,
\begin{eqnarray}
\overline{\sigma}^{reg}_{\sigma}(\omega,p/n_{l})=\frac{\sigma_{0}}{N} \sum_{r\ne r'} (n^{\sigma}_{r}-n^{\sigma}_{r'})\frac{A_{rr'}^{\sigma}}{E^{\sigma}_{r'}-E^{\sigma}_{r}}\delta(\hbar \omega -E^{\sigma}_{r}+E^{\sigma}_{r'} )
\label{optical}
\end{eqnarray}
The matrix element $A_{rr'}^{\sigma}=|\langle \Psi^{\sigma}_{r}|\widehat{j^{\sigma}_{x}}| \Psi^{\sigma}_{r'}\rangle|^{2}$, where $\widehat{j^{\sigma}_{x}}=-it\sum_{ij}(c^{\dagger}_{i,j\sigma}c_{i+\hat{x},j\sigma} -hc) $ is the $x$ component of the current operator in the $\sigma$-sector.

In Fig.\ref{Fig3} we have plotted, for $x=0.05$ and various carrier densities the regular part of the optical conductivity as a function of frequency. The calculations are realized on a $20^{3}$ sites system. A comparison with those done on both $16^{3}$ and $24^{3}$ systems has shown that the finite size effects are negligible. A systematic average over at least 200 configurations of disorder has been done. As observed experimentally, $\sigma(\omega)$ exhibits a clear broad peak for all carrier concentrations. As the hole density increases from the lowest concentration up to optimal doping $p/x=1 $ (left panel of Fig.\ref{Fig3}), we observe a clear redshift of the peak position. This feature is in agreement with the detailed experimental study performed by Burch et al. (see left panel of Fig.3 in ref. \cite{Burch06}). Moreover, this is in contrast to the valence band picture which predicts a blueshift \cite{Sinova02} (see also right panel of Fig.3 in ref. \cite{Burch06}). Note that, a blue-shift was reported in recent measurements \cite{preprint-jung10} (see Fig.1 (c) in the preprint), but no clear explanation was given to explain the origin of the conflict between earlier studies and their recent results. The data shown in Fig.1 (c) in ref.\cite{preprint-jung10} result from a fit procedure which depends on many parameters. Thus, it would be of great interest and very useful if the authors could show the measured data, this would facilitate the comparison with earlier studies. On the other hand, we have noticed that the annealed samples appear to be well annealed. Indeed, the measured Curie temperatures T$_C$ (shown in Table I of ref.\cite{preprint-jung10}) agree very well with (i) the first principle based calculations \cite{gbouzerar05a} and with those obtained in the present study (see below).

From the left panel of Fig.\ref{Fig3}, we observe that the peak is located at $0.2~t$ for the uncompensated sample ($p=x$) and approximately $0.35~t$ for the lowest concentration which corresponds to the presence of about 25$\%$ of compensating native defects as As anti-sites (double donors of electrons) for example. A direct quantitative comparison with the experimental data will be given in the next section. To our knowledge, no experimental data are available for larger hole density (beyond 1 hole/Mn), nonetheless we have also analyzed the effect of co-doping (Ga,Mn)As with a hole donor. This can be achieved experimentally by the substitution of Ga$^{3+}$ by Zn$^{2+}$ for example (Zn$^{2+}$ introduces one hole/Zn). We observe in the right panel of Fig.\ref{Fig3} that adding holes leads to an increase of the conductivity and now to a blueshift of the peak to much higher energy. For example for $p/x \ge 2.1 $, the peak is now located at approximately $1.0 t$. Such a hole concentration could be obtained in Ga$_{0.9}$Mn$_{0.05}$Zn$_{0.05}$As for example. It would be interesting to analyze experimentally the variation of $\sigma(\omega)$ as a function of y in the series of compounds Ga$_{0.95-y}$Mn$_{0.05}$Zn$_{y}$As after annealing.
Let us now discuss in more details the origin of both the redshift and blueshift obtained. For different hole density concentrations, we have been able to isolate the processes which dominate in the optical conductivity response. In other words, we have identified which matrix elements $A_{rr'}^{\sigma}$ are responsible for the peak in $\sigma^{reg}(\omega,p)$.
We could identify 3  relevant regions denoted (1),(2) and (3) depicted in Fig.\ref{Fig4}.
The origin of the red-shift is explained in the cartoon (a): below $p/x=$1, the only possible transitions are between region (1) and (2) (transitions are only allowed between occupied and empty states). The transitions between (1) and (3) are much smaller. As the hole density increases up to $p/x=1$, the Fermi energy moves down and the energy $\omega_{12}$ associated to A$_{12}$ processes reduces from 0.4~t to 0.2~t. This is the explanation of the red-shift. In the very dilute limit (typically below 1\%), this corresponds to the process between the well defined (splitted) impurity band and the top of the valence band as discussed in ref.\cite{Burch06}. For $1 \le p/x \le 2$ (case (b)) the Fermi level lies now in the region (2). The two relevant matrix elements are of type A$_{12}$ and A$_{23}$. As the hole density increases, the relative weight of A$_{23}$ increases whilst A$_{12}$ reduces. This can be seen in the right panel of Fig.\ref{Fig3} for $p/x=$ 1.35 and 1.95, where a shoulder coming from the A$_{23}$ processes is clearly visible and gets larger as the hole density increases. The typical energy associated to A$_{23}$ processes is of the order $\omega_{23} \approx t$. As we further increase the hole density (case(c)), A$_{12}$ processes are now forbidden (regions (1) and (2) are empty or filled with holes) and the only possible transitions are between the regions (2) and (3). Thus, this explains the peak located at $\frac{\omega}{t} \approx 1$ for $p/x \ge 2$ and the origin of the blue-shift.

\begin{figure}[htbp]
\includegraphics[width=6.0cm,angle=-90]{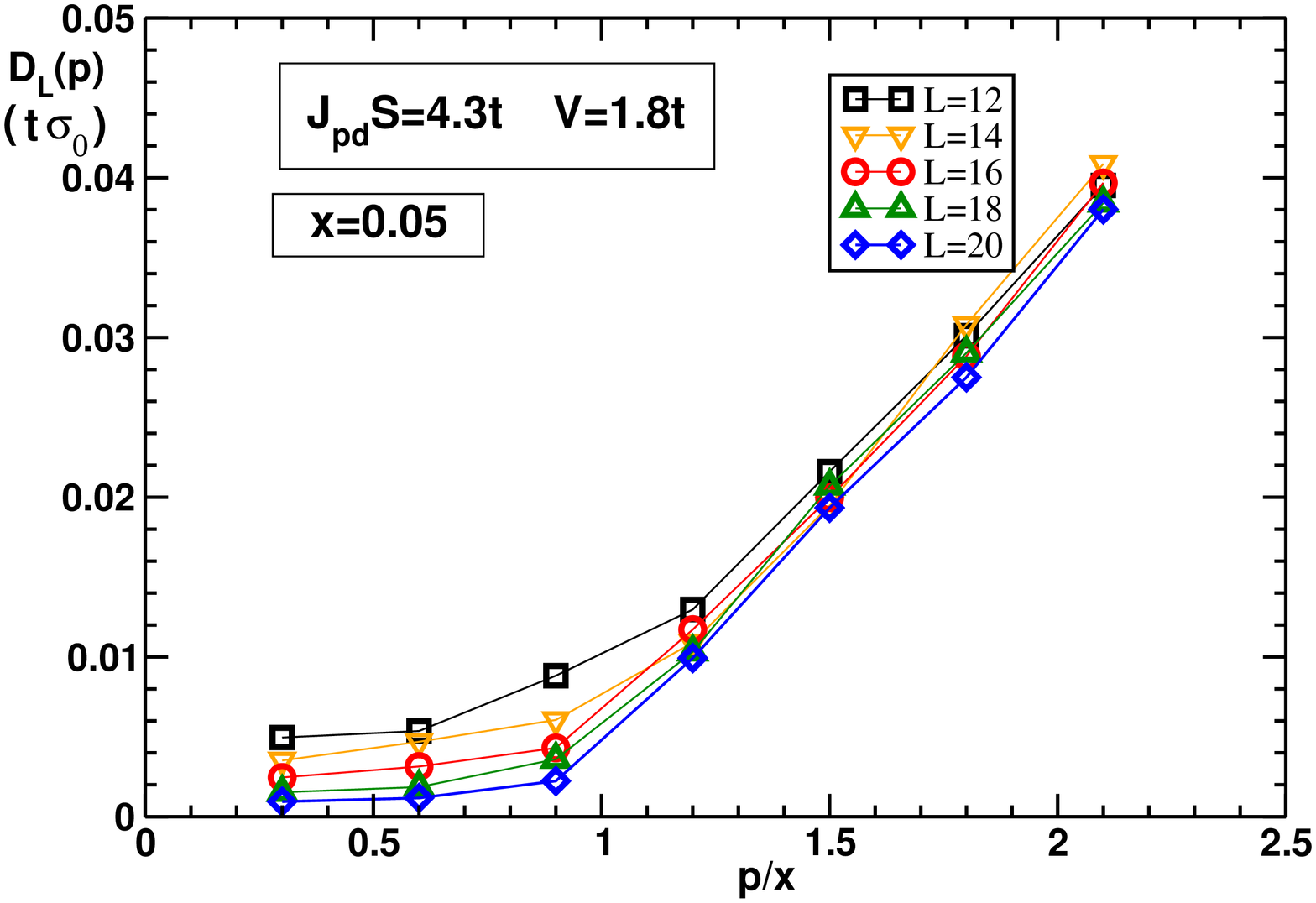}
\includegraphics[width=6.0cm,angle=-90]{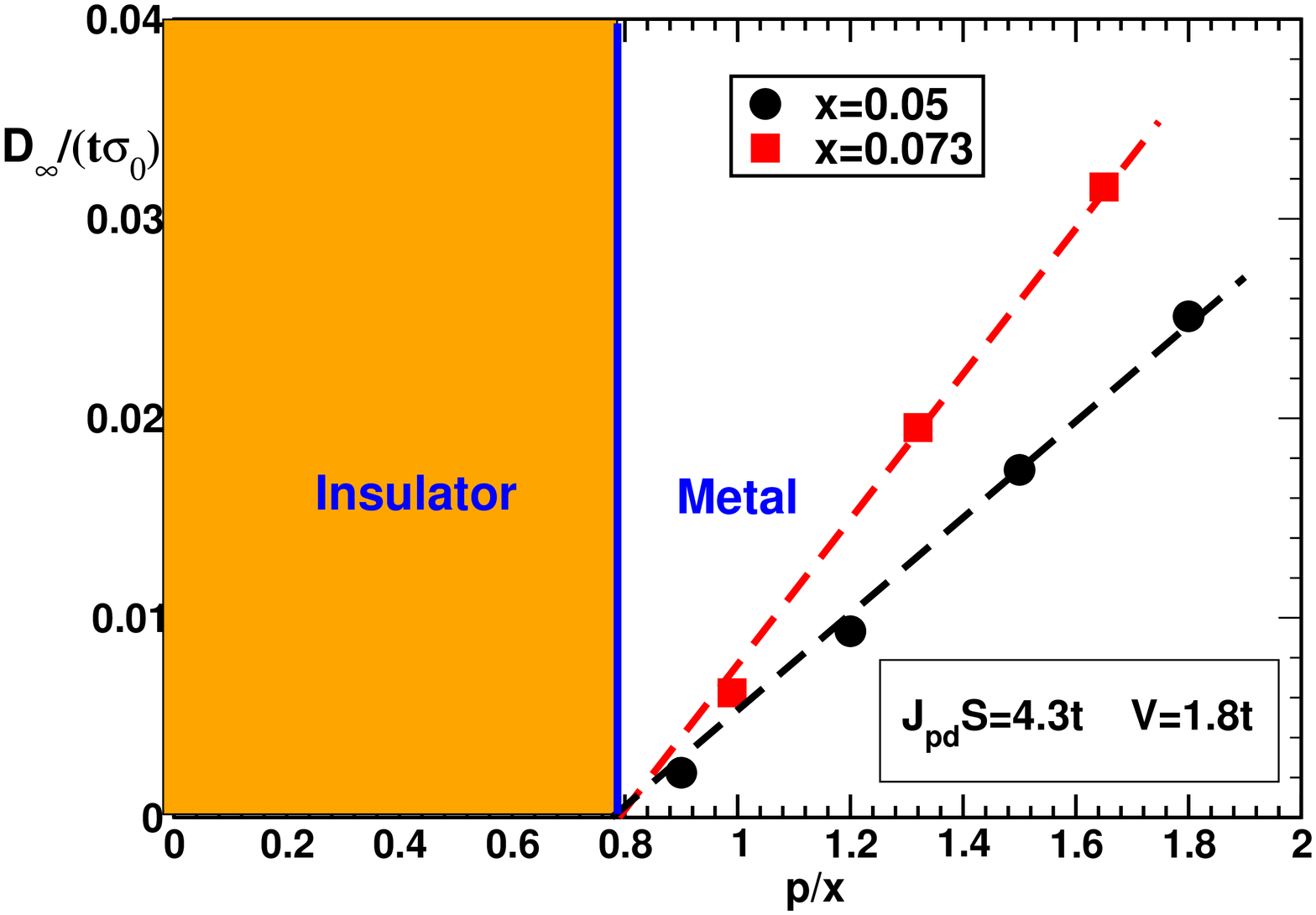}
\vspace{0.0cm}
\caption{(Color online.) (top panel) Drude weight D$_{L}$ (in units of t) as a function of the carrier density per Mn, $p/x$, calculated for various systems sizes (L=$12$,14,....20), with $x=0.05$. (bottom panel) Extrapolation in the thermodynamic limit: D$_{\infty}$, the shaded region corresponds to the insulating phase, two different Mn concentrations are shown $x=0.05$ and 0.073. The material parameters are respectively $JS=4.3t$ and $V=1.8t$}
\label{Fig5}
\end{figure}
We now proceed further and discuss the variation of the metal-insulator order parameter, the Drude weight, as a function of the carrier concentration.
Note that, the dc conductivity can be obtained from $\sigma_{DC}=D\tau$, where $\tau=l_{e}/v_{F}$, $l_{e}$ is the mean free path.
In the top panel of Fig.\ref{Fig5}, the averaged Drude weight $D_{L}$ is plotted as a function of the carrier density per Mn for various system sizes. The Mn concentration is set to $x=0.05$ and L varies from L$=$12 to 20. For the smallest system an average over 1000 configurations of disorder has been done and for the largest 200 configurations were used. After performing an extrapolation in the thermodynamic limit (see the bottom panel of fig.\ref{Fig5}), we have obtained the metal-insulator phase diagram plotted in the same figure. As seen, below a critical concentration (mobility edge) the Drude weight D is zero: the system is an insulator. Above this threshold it scales linearly with the delocalized (extended) hole concentration, $p_{\rm ex}=(p-p_{c})$.
 The transition point is located at $(\frac{p}{x})_{c} \approx 0.80$ for both Mn concentrations $x=5\%$ and $x=7.3\%$. Thus our present study predicts that the critical carrier density is $p_{c} \approx 0.0375$ for the 5$\%$ doped sample and $p_{c} \approx 0.055$ for $x=0.073$. This result is important since it indicates that approximately 20$\%$ of the carriers only contribute to the Drude weight, most of them (80$\%$) are localized. Beyond the critical concentration, the slope of the Dude weight D provides the value of the optical masses (for localized carriers $m_{\rm opt}=\infty$). Indeed, in this regime, one can write $\frac{D (p,x)}{\pi}= \frac{ p_{\rm ex} e^{2}}{m_{\rm opt}}$. We find $m_{\rm opt}(x=0.073)\approx {m_{\rm opt}(x=0.05)}\simeq 4~ m_{0}=\frac{2}{t}$, where $m_{0} \approx 0.5~m_{e}$ is the hole effective mass in GaAs host. All these results are consistent with the experimental observation that (Ga,Mn)As is close to the metal insulator phase transition and why as grown insulating samples become metallic after annealing. Note that from their experimental data Singley et al. have estimated m$_{\rm opt}$ to be within the range $0.7 m_{e} \le m_{\rm opt} \le 15 m_{e}$ \cite{Singley02}.
\begin{figure}[htbp]
\includegraphics[width=6.0cm,angle=-90]{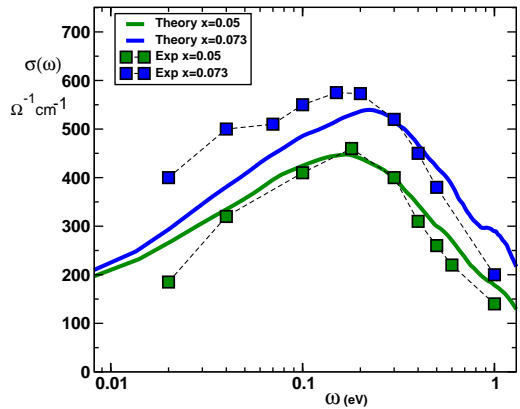}
\vspace{0.0cm}
\caption{(Color online.) Calculated and measured optical conductivity $\sigma(\omega)$ in Ga$_{1-x}$Mn$_{x}$As (in units $\Omega$$^{-1}$$\cdot$cm$^{-1}$) as a function of $\omega/t$ for the optimal doping (1hole/Mn) and for both Mn concentration $x=0.05$ and $x=0.073$.}
\label{Fig6}
\end{figure}

The natural question which arises now is the following: Is our theoretical model also able to explain the experimental data quantitatively? Let us then compare directly our calculations to the experimental data of the optical conductivity measured in annealed samples \cite{Burch06}. We will see that our model theory is able to reproduce the optical conductivity data and the Curie temperatures simultaneously. In Fig.\ref{Fig6}, we have plotted the optical conductivity in absolute units ($\Omega\cdot$cm$^{-1}$) as a function of the energy (in eV).
The experimental data for the annealed samples are taken from ref.\cite{Burch06}.
The calculations were done assuming one hole per Mn (perfectly annealed samples), taking into account, as discussed before, the fact that each Mn impurity brings three hybridized p-d states. We observe a very good and surprising quantitative agreement for both concentrations of Mn. It is even excellent in the case of the $5\%$ doped sample. The peak position is located at $200 ~meV$ for $x=0.05$ in perfect accordance with the measurements and the overall shape (peak structure, width) is also very well reproduced.
 Furthermore, using the calculated couplings $J_{i,j}(x,p)$ in the effective diluted Heisenberg Hamiltonian (see above), we have calculated within SC-LRPA the Curie temperature for both cases. We have found that T$_{C}$ is respectively 120~K and 150~K for $x=0.05$ and 0.07. These values agree very well with those measured in these annealed compounds, namely 120 and 140~K respectively \cite{Burch06}. Note also that our calculated values agree very well with those obtained using the ab-initio couplings \cite{gbouzerar05a}. It is interesting to remark that if instead of $a=\frac{a_{0}}{4^{1/3}}$ we would have used $a_{0}$, then the calculated optical conductivity would have been 60\% smaller. The small deviation between theory and experiment observed in Fig.\ref{Fig6} for $x=0.073$ may be explained as follows.
 We have used for the comparison in both cases, the total measured optical conductivity (Drude part included). As seen in Fig.1 of ref.\cite{Burch06} a small shoulder is seen at 30 meV.  The Drude contribution is larger in the $x=0.073$ compound than in the 5\% doped one and thus this may explain why below the peak, the measured optical response is a bit larger than the calculated one.

To conclude, we have calculated in a non perturbative way the optical conductivity, the phase diagram and Drude weight as a function of the hole density in GaAs samples doped with 5 $\%$ and 7.3$\%$ of Mn. In agreement with experimental measurements\cite{Burch06}, we have obtained a redshift of the optical conductivity peak.  We were also able to reproduce quantitatively the measured optical conductivity in well annealed samples. Additionally the calculated Curie temperatures are in excellent agreement with those obtained using ab initio couplings and with the experimental values. This success shows that our minimal model is able to capture not only qualitatively but also quantitatively both transport and magnetic properties in diluted magnetic semiconductors.
\\

\acknowledgments
First we would like thank E. Kats for providing us with the access to the ILL theory computer facilities and Mark Johnson for the access to the CS's clusters. We also would like to thank
 S. Kettemann, S. Srivastava and A. Chakraborty for carefully reading the manuscript and for their remarks and comments and finally we also thank
 P. Qu\'emerais, X. Blase, D. Feinberg, C. Lacroix, B. Barbara, O. C\'epas, A. Ralko for interesting and stimulating discussions.

\end{document}